\documentclass[11pt,twoside]{article}

\usepackage{asp2006}
\usepackage{epsf}
\usepackage{psfig}
\usepackage{lscape}
\usepackage{natbib}

\begin{document}
\title{Simulations of galactic disks including a dark baryonic component}   
\author{Y. Revaz\altaffilmark{1,4}, D. Pfenniger\altaffilmark{2}, F. Combes\altaffilmark{1}, F. Bournaud\altaffilmark{3}}
\altaffiltext{1}{LERMA, Observatoire de Paris}    
\altaffiltext{2}{Geneva Observatory, University of Geneva}
\altaffiltext{3}{Laboratoire AIM, CEA-Saclay DSM/Dapnia/SAp}
\altaffiltext{4}{Laboratoire d'Astrophysique, \'Ecole Polytechnique F\'ed\'erale de Lausanne (EPFL)}
 
\begin{abstract} 

$\Lambda$CDM numerical simulations predict that the ``missing baryons" reside in 
a Warm-Hot gas phase in the over-dense cosmic filaments. 
However, there are now several theoretical and observational arguments 
that support the fact that galactic disks may be more massive than usually thought, 
containing a substantial fraction of the ``missing baryons".
Hereafter, we present new N-body simulations of galactic disks, where the gas content
has been multiplied by a factor 5. The stability of the disk is ensured by assuming that the ISM 
is composed out of two partially coupled phases, a warm phase, corresponding
the observed CO and HI gas and a cold collisionless  
phase corresponding to the unseen baryons.

\end{abstract}

\vspace{-1cm}
\section{Introduction}

While the $\Lambda$CDM scenario predicts that the dark matter in galactic disks is distributed in
a dynamically hot heavy halo, there are now several observational facts showing that
a fraction of this matter may be baryonic, lying in the outer disk of galaxies. 
Long ago, \citet{bosma78} \citep[see also][]{hoekstra01} observed a correlation between the HI and dark matter in spiral
galaxies, indicating a physical link between these two components. 
This HI-DM correlation is confirmed by the baryonic Tully-Fischer relation \citep{pfenniger05}.
Dark matter in the disk of galaxies is supported by dynamical arguments based on the asymmetries of galaxies,
either in the plane of the disks or transversal to it.
Large spiral structure present in the very extended HI disk of NGC\,2915 is naturally explained by
a quasi self-gravitating disk \citep{bureau99,masset03}. Self-gravitating disks
are also vertically unstable and give a natural explanation to the old problematic of warps \citep{revaz04}.
\citet{thilker05} observed UV-bright sources in the extreme outer disk of M83, revealing unexpected star formation,
indicating that molecular gas, must be present in abundance in the outer disk of galaxies.
Other arguments supporting the idea that dark matter could reside in the galactic disk in form of
cold molecular gas are discussed in \citet{pfenniger94a}.

\section{A new multiphase ISM model}

Adding dark baryons in a galactic disk poses the problem of the global stability of the system.
This has been solved by assuming that the ISM is composed out of two dynamically decoupled phases.
A visible gas phase (vg) corresponding to the CO and HI observed gas, having velocity dispersions of about $10\,\rm{km/s}$,
and a dark gas phase (dg) of very cold, clumpy, quasi-collisionless phase \citep{pfenniger94b}. 
This latter is dynamically warmer (higher velocity dispersions) ensuring the disk stability.
The transition between the two phases depends on the gas temperature and could be computed using the
first thermodynamic principle combined with ISM heating and cooling processes. However, this method strongly 
depends on the gas density that is only a poor spatial averaged quantity in numerical simulations.
In order to avoid this density dependence, we assume that the transition times between the two phases
statistically depends simply on the local UV flux which is assumed to be the dominant ISM heating process.
%
%
The following relations for the transition times has been chosen as follow~:
	\begin{equation}
	\left\{ 
	\begin{array}{lll}
	\tau_{\rm{vg} \to \rm{dg}} &=& \tau \chi    \\
	\tau_{\rm{dg} \to \rm{vg}} &=& \tau \chi^{-1},
	\end{array}
	\right.
	\label{eq1}
	\end{equation}
where $\chi$ represents the normalized local UV flux.
The parameter $\tau$ gives the time scale of the transition, independently
of the outer flux. At equilibrium, these latter equations provide simple relations between visible gas and
dark gas density. Near young stars characterized by strong UV flux, the transition time between visible to dark gas is long
and thus the ratio visible/dark gas is high. On the contrary, in the outer galactic regions dominated by a weak extragalactic
UV flux, the transition between visible to dark gas is very short and the dark gas dominates there. 

\vspace{-0.25cm}
\section{Application to isolated disks}

In this section, we compare the evolution of two models of isolated spiral galaxies different
by their dark matter content. The first model is a ``classical" galaxy having a collisionless massive dark halo 
and no additional dark baryons. In the second model, a fraction of the dark halo mass is transferred into a 
dark baryonic gaseous disk (the baryonic gas is multiplied by a factor of 5) that takes part to the ISM cycle described earlier.
The galaxy is constructed in order to share the same observational properties than the ``classical" model and being consistent
with Eq.~\ref{eq1}:
a similar stellar component (exponential disk + bulge), a similar rotation curve (flat up to 40\,\rm{kpc}), 
and a similar surface density of the visible gas (up to $20\,\rm{kpc}$). The dark disk is as large as
the dark matter halo, extending up to $90\,\rm{kpc}$. Its contribution to the total surface density is negligible 
at the center, while it is large in the outer part, where the UV field is weak.
\begin{figure}[!ht]
\plotone{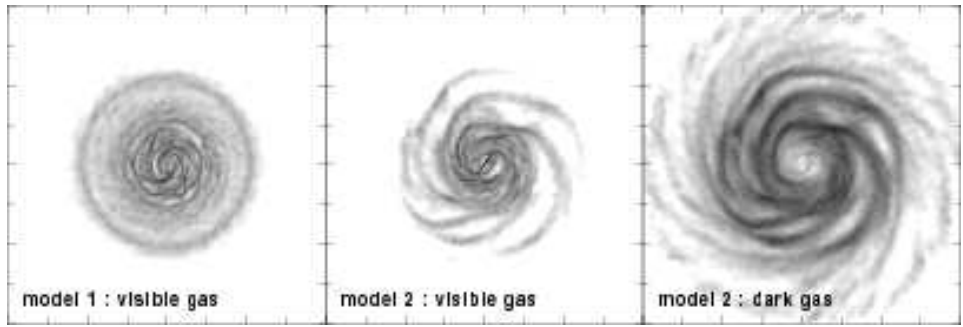}
\caption{Comparison between the surface density of the visible gas of model 1 and visible and dark gas of model 2,
after $2\,\rm{Gyr}$ of evolution. The size of each box is $200\times200\,\rm{kpc}$.}
\label{fig1}
\end{figure}
%
The initial conditions have been computed by resolving the second moments of the Jeans equation. The radial velocity dispersion
of the visible gas is set to $10\,\rm{km/s}$ while for the dark gas, it is set such that the Safronov-Toomre
parameter $Q$ to 1, ensuring disk stability.
The models have been run using a modified version of the Gadget-2 code \citep{springel05}, 
including the ISM cycling. Model 1 and 2 have respectively  $524'285$ and $804'751$ particles ensuring a 
spatial resolution of $250\,\rm{pc}$.

The evolution of the disks  shows a similar global behavior between the two models.
After $4\,\rm{Gyr}$ both models have a very similar rotation curve and both have formed a comparable bar.
Despite having multiplied the baryonic gaseous disk by a factor 5, model 2 is globally stable.
It's stability is ensured by the larger velocity dispersion of the collisionless dark gas, being about twice the one
of the visible gas. 
Fig.~\ref{fig1} compares the surface density of the visible disk after $2\,\rm{Gyr}$. 
While the azimuthal averaged surface density is similar between the two models, the spiral structure 
in the outer $20\,\rm{kpc}$ is very different. Model 1 has nearly no spiral structure while model 2 displays open 
and contrasted spiral arms up to the edge of the visible disk.
The presence of the dark gas in model 2 increases the surface density and makes the Safronov-Toomre 
parameter\footnote{Here, the Safronov-Toomre parameter is computed by taking into account only the disk component : 
stars, visible and dark gas, for the surface density and radial velocity dispersion estimation.} be constant around a value of 2. The self-gravity of the disk is increased making it quasi-stable. 
On the contrary, in model 1, the surface density is much smaller and Q is about twice larger, ensuring the stability of the disk. 
This behavior has to be related with puzzling observations of spiral structures in HI disk, difficult to be explained by
the self-gravity of the HI disk itself \citep{bureau99}.

\vspace{-0.25cm}
\section{Application to the missing mass in tidal dwarf problem}

\begin{figure}[!ht]
\plotone{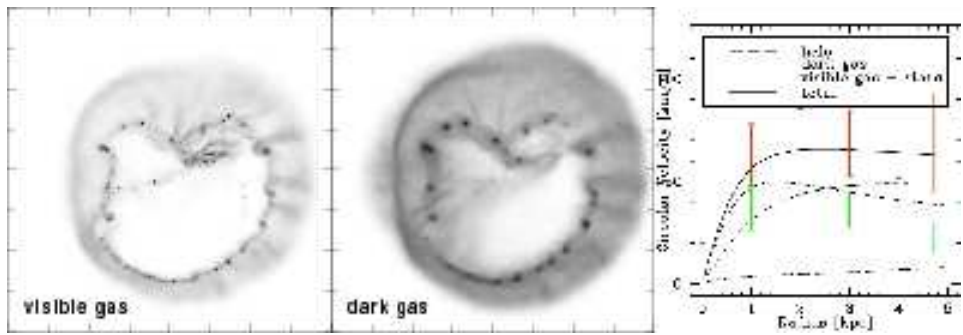}
\caption{Left: surface density of the visible gas, 
Middle: surface density of the dark gas. The size of each box is $200\times200\,\rm{kpc}$, 
Right: Velocity curve of a typical tidal dwarf galaxy, compared with data of \citet{bournaud07}.}
\label{fig2}
\end{figure}

Observation of the system NGC5291 revealed that the tidal dwarf galaxies present in the debris of NGC5291 have 
an unexpected dark matter content.
In this section, we show that the presence of dark gas in the debris of galaxies is a natural product of the
assumption that galaxies have an additional dark gas component.
Using the initial condition of \citet{bournaud07}, we have reproduced the HI ring of NGC5291 using our new
multiphase model. The initial spiral galaxy is pretty similar to the model 2 (see below), with a slightly more massive disk. 
The left and middle panel of Fig.~\ref{fig2} show the surface density of the gas, $400\,\rm{Myr}$ after the collision
in a box of $200\times200\,\rm{kpc}$. 
Tidal dwarf galaxies have formed in a ring, out of the debris of the dissipative visible gas. 
The low velocity dispersion of the dark gas (compared to the dynamically hot dark halo) lets it fall into potential 
wells of the tidal dwarfs, adding a dark component to this latter objects. 
The typical rotation curve of the tidal dwarfs formed is display on the right panel
of Fig.~\ref{fig2}. In these objects, the dark gas (dotted curve) plays the same role than a halo in a spiral galaxy. 
It allows the total rotation curve (plain curve) to stay straight up to $6\,\rm{kpc}$, while the contribution of the visible
baryons (star + gas) decreases. Nearly no dark halo mass is contained in these objects and its contribution 
to the total rotation curve (dashed-dotted curve) is negligible.
In this latter figure, the upper error bars correspond to the observed velocity curve of NGC5291a
and the latter to the modeled velocity curve of its  visible component \citep{bournaud07}. A good
agreement between model and observations is obtained.
\vspace{-0.75cm}

\section{Conclusion}

New numerical simulations show that the hypothesis where galaxies are assumed to have
an additional dark baryonic component is not in contradiction with galaxy properties.
Multiplying the HI disk by a factor of 5 let the disk stable
if the ISM is composed out of two dynamically decoupled phases.
Such galaxy gives a natural solution to different puzzling problems like the
the presence of spiral structures in the outer HI disks and the missing mass problem in
collisional debris of galaxies.


\begin{thebibliography}{1}
\expandafter\ifx\csname natexlab\endcsname\relax\def\natexlab#1{#1}\fi

\bibitem[{{Pfenniger} \& {Combes}(1994)}]{pfenniger94b}
{Pfenniger}, D. \& {Combes}, F. 1994, \aap, 285, 94

\end{thebibliography}


\vspace{-0.25cm}
\begin{thebibliography}{}


\bibitem[{{Bosma}(1978)}]{bosma78}
{Bosma}, A. 1978, PhD Thesis, Groningen Univ., (1978)


\bibitem[{{Bournaud} {et~al.}(2007){Bournaud}, {Duc}, {Brinks}, {Boquien},
  {Amram}, {Lisenfeld}, {Koribalski}, {Walter}, \& {Charmandaris}}]{bournaud07}
{Bournaud}, F., {Duc}, P.-A., {Brinks}, E., {et~al.} 2007, Science, 316, 1166


\bibitem[{{Bureau} {et~al.}(1999){Bureau}, {Freeman}, {Pfitzner}, \&
  {Meurer}}]{bureau99}
{Bureau}, M., {Freeman}, K.~C., {Pfitzner}, D.~W., \& {Meurer}, G.~R. 1999,
  \aj, 118, 2158


\bibitem[{{Hoekstra} {et~al.}(2001){Hoekstra}, {van Albada}, \&
  {Sancisi}}]{hoekstra01}
{Hoekstra}, H., {van Albada}, T.~S., \& {Sancisi}, R. 2001, \mnras, 323, 453

\bibitem[{{Masset} \& {Bureau}(2003)}]{masset03}
{Masset}, F.~S. \& {Bureau}, M. 2003, \apj, 586, 152

\bibitem[{{Pfenniger} {et~al.}(1994){Pfenniger}, {Combes}, \&
  {Martinet}}]{pfenniger94a}
{Pfenniger}, D., {Combes}, F., \& {Martinet}, L. 1994, \aap, 285, 79

\bibitem[{{Pfenniger} \& {Combes}(1994)}]{pfenniger94b}
{Pfenniger}, D. \& {Combes}, F. 1994, \aap, 285, 94

\bibitem[{{Pfenniger} \& {Revaz}(2005)}]{pfenniger05}
{Pfenniger}, D. \& {Revaz}, Y. 2005, \aap, 431, 511

\bibitem[{{Revaz} \& {Pfenniger}(2004)}]{revaz04}
{Revaz}, Y. \& {Pfenniger}, D. 2004, \aap, 425, 67

\bibitem[{{Springel}(2005)}]{springel05}
{Springel}, V. 2005, \mnras, 364, 1105

\bibitem[{{Thilker} {et~al.}(2005){Thilker}, {Bianchi}, {Boissier}, {Gil de
  Paz}, {Madore}, {Martin}, {Meurer}, {Neff}, {Rich}, {Schiminovich},
  {Seibert}, {Wyder}, {Barlow}, {Byun}, {Donas}, {Forster}, {Friedman},
  {Heckman}, {Jelinsky}, {Lee}, {Malina}, {Milliard}, {Morrissey}, {Siegmund},
  {Small}, {Szalay}, \& {Welsh}}]{thilker05}
{Thilker}, D.~A., {Bianchi}, L., {Boissier}, S., {et~al.} 2005, \apjl, 619, L79

\end{thebibliography}
\end{document}